\documentclass{article}

\usepackage{arxiv}

\usepackage[utf8]{inputenc} 
\usepackage[T1]{fontenc}    
\usepackage{hyperref}       
\usepackage{url}            
\usepackage{booktabs}       
\usepackage{amsfonts}       
\usepackage{nicefrac}       
\usepackage{microtype}      
\usepackage{lipsum}		
\usepackage{graphicx}
\usepackage{natbib}
\usepackage{doi}
\usepackage{tabularx}
\usepackage{array}
\usepackage{amsmath}
\usepackage{orcidlink}

\title{Practical Evaluation of Quantum Kernel Methods for Radar Micro-Doppler Classification on Noisy Intermediate-Scale Quantum (NISQ) Hardware}

\author{
	Dr.~Vikas~Agnihotri\orcidlink{0009-0007-3284-3423} \\
	Independent Researcher \\
	\texttt{er.vikasagnihotri@gmail.com}
	\and
	Jasleen~Kaur \\
	National Institute of Technology, Rourkela \\
	\texttt{jasstronger@gmail.com}
	\and
	Sarvagya~Kaushik \\
	Indian Institute of Technology, Dhanbad
}

\renewcommand{\shorttitle}{\textit{arXiv} Template}

\hypersetup{
pdftitle={Exploiting Micro-Doppler Signatures for Radar Classification: A Quantum SVM Approach},
pdfsubject={q-bio.NC, q-bio.QM},
pdfauthor={David S.~Hippocampus, Elias D.~Striatum},
pdfkeywords={First keyword, Second keyword, More},
}

\begin{document}
\maketitle 

\begin{abstract}
	
    This paper examines the application of a Quantum Support Vector Machine (QSVM) for radar-based aerial target classification using micro-Doppler signatures. Classical features are extracted and reduced via Principal Component Analysis (PCA) to enable efficient quantum encoding. The reduced feature vectors are embedded into a quantum kernel-induced feature space using a fully entangled ZZFeatureMap and classified using a kernel based QSVM. Performance is first evaluated on a quantum simulator and subsequently validated on NISQ-era superconducting quantum hardware, specifically the IBM Torino (133-qubit) and IBM Fez (156-qubit) processors. Experimental results demonstrate that the QSVM achieves competitive classification performance relative to classical SVM baselines while operating on substantially reduced feature dimensionality. Hardware experiments reveal the impact of noise and decoherence and measurement shot count on quantum kernel estimation, and further show improved stability and fidelity on newer Heron r2 architecture. This study provides a systematic comparison between simulator-based and hardware-based QSVM implementations and highlights both the feasibility and current limitations of deploying quantum kernel methods for practical radar signal classification tasks. 
    
\end{abstract}

\keywords {Quantum Support Vector Machine (QSVM) \and Quantum Kernel Methods \and Radar Micro-Doppler Signatures \and Aerial Target Classification \and Principal Component Analysis (PCA) \and NISQ Quantum Computing \and Superconducting Quantum Hardware \and IBM Quantum \and Quantum Machine Learning for Radar \and Target Classification.}

\section{Introduction}

The Doppler effect causes a shift in observed frequency due to relative motion between a source and an observer \cite{sasso2020doppler}. Rapid inherent secondary target motions, such as vibration or rotation of a helicopter rotor blade, generate Micro-Doppler signatures. These distinct frequency modulations reveal fine-grained target dynamics beyond just position and velocity \cite{peter2021microdoppler}. Because these fast-spinning rotor blades of the helicopter can produce extreme and distinct micro-Doppler shifts, the resulting signature is so specific that it can identify the type of helicopter or drone, and sometimes even the number of blades and their rotation speeds \cite{chen2006microdoppler}.Consequently, distinguishing between these targets is a significant challenge, as the spectral features are often subtle, overlapping, and exhibit inherent nonlinearity that classical linear classifiers find difficult to resolve. 

To address these issues, this study explores the use of quantum-enhanced classifiers. In practice, classical SVMs face significant computational bottlenecks during kernel estimation when the feature dimensionality increases. In this work, a quantum kernel–based SVM is implemented using the ZZFeature Map, an entangling feature map. PCA is used to reduce the classical feature dimensionality prior to encoding into a quantum feature space induced by an entangling ZZFeatureMap, a complex vector space where the dimensionality increases exponentially with the number of qubits  ($2^n$) \cite{havlicek2019supervised}. By mapping the data into this vastly expanded quantum state space, micro-Doppler patterns can become more easily separable \cite{schuld2019quantum}. Then it allows the classifier to construct an optimal separating hyperplane within the high-dimensional Hilbert space, effectively resolving computationally expensive non-linearities of classical kernel methods \cite{biamonte2017quantum}.

To evaluate the proposed approach, this work provides a systematic comparison between the following implementations:
\begin{itemize}
    \item Classical Baseline: SVM utilizing a Radial Basis Function (RBF) kernel with the full 15-dimensional feature set.
    \item Quantum Simulator \cite{havlicek2019supervised,qiskitQSVM}: QSVM utilizing a $ZZFeatureMap$ with PCA-reduced features to establish ideal performance benchmarks.
    \item Physical Hardware (133-qubit \cite{ibmTorino}): Validation on the IBM Torino processor to assess performance under realistic noise conditions.
    \item Physical Hardware (156-qubit \cite{ibmFez}): Validation on the IBM Fez processor to evaluate the impact of architectural improvements on classification fidelity.
\end{itemize}

\section{Literature Survey}
Although analysing the micro-Doppler effect and target classification on radars is challenging, a substantial body of work has explored classical and quantum approaches. Support Vector Machines (SVM) have been proven to be an important method for classification. Early studies showed that SVM has been efficient on micro-Doppler signatures for small movements in human activity and mechanical targets. Raw radar signals are converted into spectrograms to bring out particular dynamical and statistical variables, which are further classified by an SVM using a kernel-based decision boundary.Classical machine learning techniques, particularly Support Vector Machines (SVMs), have been widely applied to micro-Doppler–based radar classification.  Early work by Sigg et. al. \cite{sigg2015support} demonstrated that kernel-based SVMs could effectively discriminate between humans and vehicles using micro-Doppler signatures, achieving high accuracy under controlled conditions. Ritchie et al. \cite{ritchie2017multistatic} further exploited harmonic spacing in micro-Doppler spectra to distinguish UAVs from birds and other aerial targets, while Zhang et al.  \cite{zhang2017vehicle} employed time–frequency representations such as the Short-Time Fourier Transform (STFT) to classify ground vehicles based on rotating mechanical components.

Despite their effectiveness, these classical approaches remain sensitive to noise, clutter, and limited training data. To address these limitations, subsequent studies introduced statistical noise and clutter modeling frameworks for radar environments with non-Gaussian characteristics \cite{noise2019statistical} as well as transfer learning techniques that adapt pre-trained deep neural networks to micro-Doppler datasets  \cite{smith2024transfer}.  While these methods improve robustness under adverse conditions, they rely heavily on large-scale classical models and do not fundamentally address the computational bottlenecks associated with high-dimensional kernel methods. Another work \cite{podgorski2024improved} proposed the use of quantum feature space along with optimization, reducing reliance on large datasets. \cite{suzuki2024quantum} was the first to demonstrate a theoretical leap by showing that QVSMs successfully classify and regress on physical trapped-ion qubits, but only on low-dimensional feature sets.

Quantum machine learning (QML) has emerged as a promising approach for learning in high-dimensional feature spaces by encoding classical data into quantum states. An early and influential overview by Biamonte et al. \cite{biamonte2017quantum} surveyed the landscape of quantum-enhanced learning, outlining key concepts such as quantum data encoding, quantum linear algebra, and hybrid learning architectures. Schuld and Killoran \cite{schuld2019quantum} subsequently provided a rigorous mathematical treatment of quantum feature maps, showing how classical data can be embedded into Hilbert spaces where kernel-based learning may offer advantages over classical counterparts. Building on these theoretical developments, Havlíček et al. \cite{havlicek2019supervised} introduced the concept of quantum kernel estimation (QKE) and demonstrated supervised classification using entangling feature maps on superconducting quantum processors, marking one of the first experimental validations of quantum-enhanced kernel methods.

Subsequent research investigated parameterized quantum circuits (PQCs) as an alternative to kernel-based learning, in which trainable parameters are embedded directly within the quantum circuit structure \cite{schuld2020circuit}. These models offer considerable expressive power, but in practice their performance is limited by increasing circuit depth, optimization instability during training, and sensitivity to hardware noise. In parallel, the Qiskit framework \cite{qiskit2021framework} enabled accessible implementations of QSVMs on both quantum simulators and real superconducting hardware. Nevertheless, most reported studies remain restricted to low-dimensional feature representations and do not provide a systematic analysis of the effects of measurement shot count, device noise, or hardware-generation differences on classification performance.
 
 Moving from theory to physical implementation necessitates overcoming hardware instability, Temme \cite{temme2017error} introduced critical error-mitigation protocols for "short-depth" circuits, enabling Quantum SVMs to maintain accuracy on NISQ devices. However, current literature still lacks quantified data on how environmental factors, specifically like phase noise and multi-path interference degrade these quantum states, establishing a clear need for the robust noise modeling proposed in our research.

 Research since 2023 has rapidly integrated advanced learning paradigms into radar signal processing, moving from classical transfer learning to hybrid quantum architectures. Addressing the challenge of limited training data, \cite{smith2024transfer} successfully adapted pre-trained deep feature extractors to radar, achieving 96.1\% accuracy on helicopter datasets even under varying noise levels. Simultaneously, the focus shifted toward QML to enhance classification margins; Lee \cite{podgorski2024improved} demonstrated that "entangling" quantum feature maps could improve the Area Under Curve (AUC) by 2\% in simulated environments. Moving from simulation to hardware, \cite{suzuki2024quantum} validated these benefits on ion-trap devices, reporting a 1.8\% accuracy gain over classical methods. Finally, Zhao \cite{wang2024hqnn} combined these approaches in a hybrid quantum-classical pipeline, which improved robustness against multipath interference by 3\%.

\section{Mathematical Formulation}
\subsection{Classical SVM}
The classical soft-margin SVM formulation follows the standard framework introduced by Cortes and Vapnik \cite{cortes1995svm}, with kernelized extensions as described by Schölkopf and Smola \cite{scholkopf2002learning}.

Given training data $\{(x_i, y_i)\}_{i=1}^m$, where $y_i \in \{+1, -1\}$, the Support Vector Machine (SVM) solves the following optimization problem:

\[
\min_{w, b, \xi} \ \frac{1}{2} \|w\|^2 + C \sum_{i=1}^m \xi_i
\]
subject to
\[
y_i \left( w^T \phi(x_i) + b \right) \ge 1 - \xi_i, \quad \xi_i \ge 0, \quad i = 1, \dots, m
\]

where $\phi(x)$ is an implicit feature mapping defined by a kernel function
\[
K(x, x') = \langle \phi(x), \phi(x') \rangle.
\]

For the Radial Basis Function (RBF) kernel, we have
\[
K(x, x') = \exp\left( -\gamma \|x - x'\|^2 \right),
\]
where $\gamma > 0$ is a kernel parameter.

\subsection{Quantum SVM}

Quantum Support Vector Machines (QSVMs) extend classical kernel-based learning by encoding input data into quantum states through a parameterized feature map. Given a classical input vector  $\boldsymbol{x}$, the corresponding quantum feature state is prepared using a unitary transformation $U_{\boldsymbol{x}}$ acting on an $n$-qubit reference state \cite{schuld2019quantum}:

\[
|\phi(\boldsymbol{x})\rangle = U_{\boldsymbol{x}} |0^n\rangle.
\]

The similarity between two data points is quantified through a quantum kernel defined by the squared inner product of their corresponding quantum states, which can be estimated using quantum circuits. The resulting kernel matrix is then supplied to a classical SVM solver for training and classification.

The \textit{quantum kernel} between two inputs $\boldsymbol{x}$ and $\boldsymbol{x'}$ is defined as:

\[
K_q(\boldsymbol{x}, \boldsymbol{x'}) = \left| \langle \phi(\boldsymbol{x}) | \phi(\boldsymbol{x'}) \rangle \right|^2.
\]

We employ the following parameterized quantum circuit for the feature map:

\[
U_{\boldsymbol{x}} = \left( \prod_{j=1}^{n} H_j \right) \left( \prod_{j=1}^{n} R_Z(x_j) \right) \left( \prod_{j=1}^{n} H_j \right),
\]
constructed using PCA-reduced feature vectors onto $n$ qubits. The computed kernel values $K_q(\boldsymbol{x}, \boldsymbol{x'})$ are then supplied to a classical SVM solver for classification.

\section{Methodology}

\begin{figure*}[!t]
    \centering
    \includegraphics[width=\linewidth]{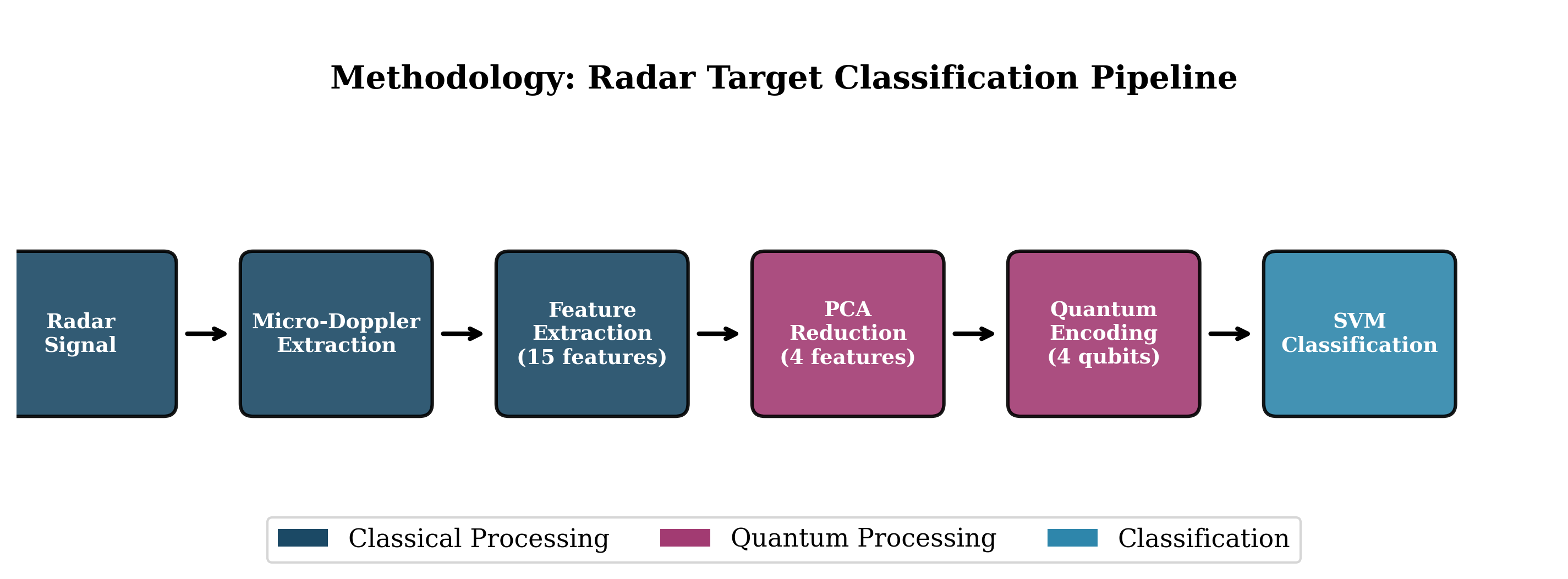}
    \caption{Radar classification pipeline with classical and quantum SVM}
    \label{fig:radar_classi_pipeline}
\end{figure*}

The methodology (Figure \ref{fig:radar_classi_pipeline}) follows a sequential flow by performing feature extraction on 15 different features from micro-Doppler signatures in raw radar signals, reducing them via PCA, encoding them, and then classifying targets using SVM. 

\subsection{Radar Simulation \& Data Preparation \cite{agnihotri2019frequency,agnihotri2020automatic}:}
To generate data for evaluating the SVM and QSVM classifiers, a simulated 9~GHz (X-band) radar system was employed with the following parameters:
\begin{itemize}
    \item Carrier frequency: 9~GHz
    \item Wavelength: 0.033~m
    \item Sampling frequency: 10~kHz
    \item Signal duration: 0.5~s
    \item Signal-to-noise ratio (SNR): 5--15~dB
\end{itemize}

\begin{figure*}[!t]
    \centering
    \includegraphics[width=\linewidth]{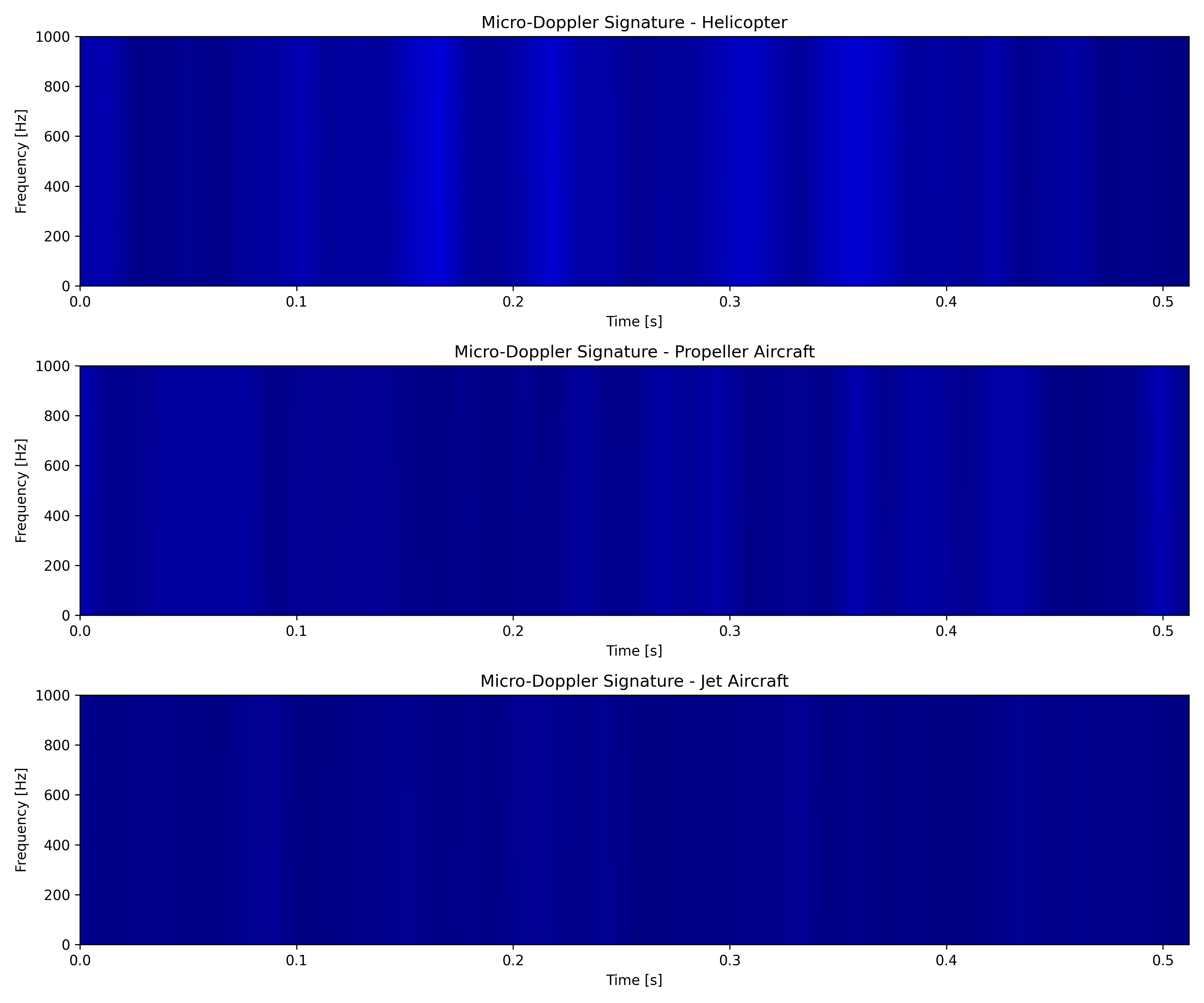}
    \caption{Micro-Doppler Signatures of Helicopter, Propeller Aircraft and Jet Aircraft.}
    \label{fig:micro- doppler signatures}
\end{figure*}

In the context of classification task ( Figure \ref{fig:micro- doppler signatures}), three distinct aerial target classes were simulated:
\begin{itemize}
    \item Helicopter class is identified by pronounced rotor blade modulation, characterized by the presence of 3-5 rotor blades, a rotation frequency ranging from 15 to 25 Hz, and blade lengths between 2.5 and 4.0 meters [6].
    \item Propeller Aircraft category is characterized by moderate blade flashes. They typically have two propellers rotating at a frequency of 50 Hz with a propeller radius of 1.5 meters.
    \item Jet Aircraft class is characterized by smooth Doppler signatures. Their engine modulates at frequencies between 80 and 120 Hz, producing high-velocity body returns.
\end{itemize}

Realistic atmospheric effects were simulated to include: 
\begin{itemize}
    \item Weather conditions to include - clear sky conditions with no attenuation, light rain with an attenuation range from 0.5 to 2 dB/km, heavy rain with an attenuation range from 5 to 15 dB/km and fog with an attenuation range from 0.1 to 0.5 dB/km.
    \item Turbulent atmospheric conditions are characterized by fluctuating phase and amplitude.
\end{itemize}

\subsection{Feature Extraction:}

Using the Short-Time Fourier Transform (STFT), a total of 15 discriminative micro-Doppler features are extracted from back scattered signal as described in \cite{agnihotri2019features}. The STFT spectrogram provides a concise representation of the non-stationary micro-motion nature of the received signal in the time–frequency domain. The features are organized as follows:

\begin{figure*}[!t]
    \centering
    \includegraphics[width=\linewidth]{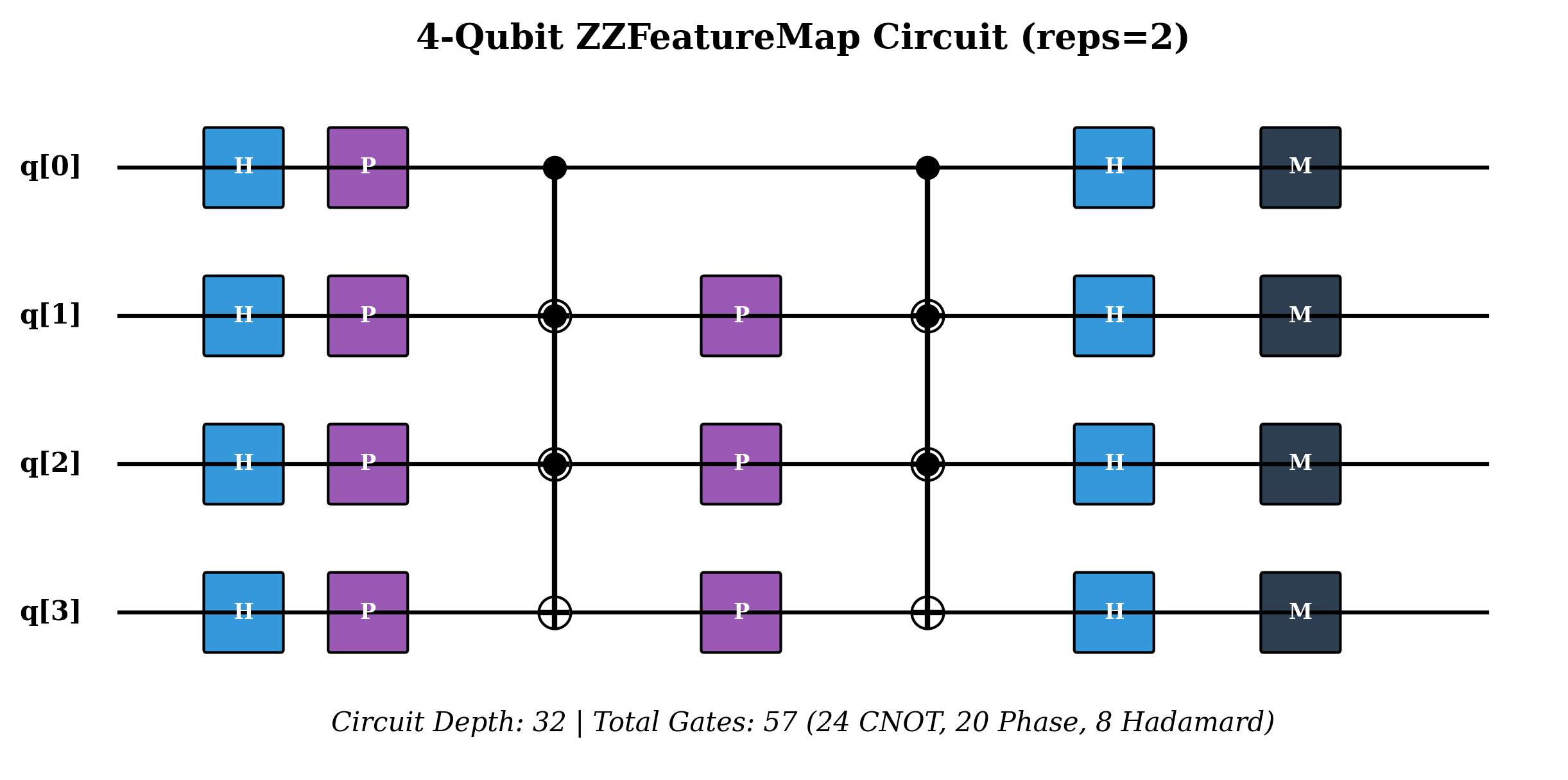}
    \caption{4-qubit ZZFeatureMap quantum circuit with two repetitions used for feature encoding.}
    \label{fig3_quantum_circuit}
\end{figure*}

\begin{itemize}
    \item \textbf{Statistical moments:} mean, standard deviation, skewness and kurtosis.
    \item \textbf{Spectral characteristics:} bandwidth and peak frequency.
    \item \textbf{Temporal dynamics:} mean, variance and maximum change.
    \item \textbf{Spectral properties:} entropy and centroid.
    \item \textbf{Spectral transitions:} roll off and flatness.
    \item \textbf{Dominant frequency ratios:} relative strength of the principal spectral components.
\end{itemize}

These features create a compact 15-dimensional representation. This captures the essential spectral, temporal, and dynamical characteristics of micro-Doppler signatures, enabling robust classification.

\subsection{Classical SVM Setup:}  
The study implements a Classical SVM with a Radial Basis Function (RBF) kernel to effectively capture the non-linear separability present in micro-Doppler features. The regularization parameter \(C \) is set to 10, achieving a balance between maximizing the margin and minimizing the classification error. The kernel coefficient \(\gamma\) is chosen using the \textit{scale} heuristic, which adjusts to the input data's variance. All 15 extracted micro-Doppler features served as inputs. Applying Z-score normalisation to each feature ensures a zero mean and unit variance. This improves numerical stability and convergence during training.

\subsection{QSVM Setup:}  

\begin{table}[htbp]
\centering
\caption{Quantum Circuit Properties (ZZFeatureMap)}
\label{tab:circuit}
\begin{tabular}{cc}
\toprule
\textbf{Property} & \textbf{Value} \\
\midrule
Number of Qubits        & 4 \\
Classical Bits          & 4 \\
Circuit Depth           & 32 \\
Feature Map             & ZZFeatureMap \\
Entanglement            & Full \\
Repetitions (reps)      & 2 \\
\midrule
\multicolumn{2}{l}{\textbf{Gate Count}} \\
\midrule
Hadamard (H)            & 8 \\
Phase (P)               & 20 \\
CNOT (CX)               & 24 \\
Measurement             & 4 \\
\midrule
\textbf{Total Gates}    & 57 \\
\bottomrule
\end{tabular}
\end{table}

The QSVM is configured to utilize quantum feature mapping and kernel estimation, improving the non-linear separability of micro-Doppler features. The overall configuration involves feature reduction using PCA, a quantum feature map based on the ZZFeatureMap, and a fully entangled circuit with properties specified in Table \ref{tab:circuit}. Figure \ref{fig3_quantum_circuit} represents the quantum circuit of a 4-qubit ZZFeatureMap with two repetitions, illustrating feature encoding through parameterized phase rotations and entangling CNOT operations prior to measurement. 

\subsubsection{Feature Reduction (PCA and Qubit Selection):}  
The number of qubits, which corresponds to the reduced feature dimensionality, is empirically determined using a PCA sweep from 2 to 12 components. This analysis finds the ideal balance between variance retention, classification performance, and hardware feasibility. Table~\ref{tab:pca_sweep} summarizes the results.

\newcolumntype{C}{>{\centering\arraybackslash}X}

\begin{table}[h]
\centering
\caption{Feature Dimensionality Sweep Analysis}
\label{tab:pca_sweep}
\small
\renewcommand{\arraystretch}{1.15}
\begin{tabular}{
		>{\centering\arraybackslash}p{1.7cm}
		>{\centering\arraybackslash}p{2.5cm}
		>{\centering\arraybackslash}p{2cm}
		>{\centering\arraybackslash}p{2.2cm}
	}
\hline
\textbf{Qubits} & \textbf{PCA Variance (\%)} & \textbf{Accuracy (\%)} & \textbf{$\Delta$ Accuracy (\%)} \\
\hline
2 & 61.30 & 89.63 & -- \\
3 & 71.59 & 93.33 & +3.70 \\
4 & 81.58 & 94.81 & +1.48 \\
5 & 87.95 & 94.81 & +0.00 \\
6 & 94.00 & 91.85 & $-$2.96 \\
\hline
\end{tabular}
\end{table}

Following this sweep, four qubits are chosen based upon the specified criteria:
\begin{itemize}
    \item \textbf{Variance threshold:} Configuration using four qubits is the first configuration exceeding the 80\% explained variance criterion (81.58\%).
    \item \textbf{Elbow point:} The accuracy improvement plateaus after four qubits and remains constant from 4 to 5 qubits.
    \item \textbf{Overfitting prevention:} Accuracy drops beyond five qubits suggesting susceptibility to noise and overfitting.
    \item \textbf{NISQ compatibility:} Four-qubit circuits offer shallow-depth implementations, suitable for NISQ hardware. This makes them ideal for reliable performance on IBM quantum devices.
\end{itemize}

Therefore, four qubits provide an optimal balance between expressive power, generalization performance, and hardware feasibility.

\subsubsection{Quantum Feature Map (ZZFeatureMap):}  
The QSVM uses the \textit{ZZFeatureMap} to transform classical data into quantum states. This feature map employs Pauli-Z operators \cite{havlicek2019supervised} and controlled phase interactions to capture pairwise correlations between features. The classical data vector $\mathbf{x}$ is mapped into a quantum state using:
\[
|\phi(\mathbf{x})\rangle = U_{\Phi}(\mathbf{x}) |0\rangle^{\otimes n},
\]
where $n$ is the number of qubits and $U_{\Phi}(\mathbf{x})$ represents the parameterized unitary transformation given by:
\[
U_{\Phi}(\mathbf{x}) = \exp\left(i \sum_{i<j} \phi_{ij}(\mathbf{x}) Z_i Z_j \right) \cdot H^{\otimes n}.
\]
where, $H$ represents the Hadamard gate and $Z_i Z_j$ are Pauli-Z interaction terms that introduce entanglement-dependent phase shifts. This permits a non-linear quantum embedding of the classical features.

\subsubsection{Circuit Design and Entanglement:}  
The quantum circuit is built with full entanglement connectivity, enabling every qubit to interact with every other one. This maximizes the expressive capacity of the quantum feature map by allowing complex multi-qubit correlations. These are crucial for modeling the highly non-linear structure of micro-Doppler signatures.

\subsubsection{Hilbert Space Representation:}  
The 4-qubit system corresponds to a 16-dimensional complex Hilbert space, enabling richer nonlinear embeddings compared to the original PCA space. This exponential expansion of the feature space allows QSVM to achieve improved separability of complex patterns that are challenging to resolve in classical feature spaces.  It also maintains a circuit depth compatible with current NISQ-era quantum hardware.

Overall, this QSVM configuration implements a principled and hardware-aware framework. This framework balances feature representation power, generalization capability, and practical implement-ability on a real quantum processors.

\subsection{Dataset and Quantum Hardware:}

The dataset contains 450 samples, with each class containing 150 samples. A 70/30 train–test split is employed to ensure a balanced evaluation of model performance. To guarantee reproducibility, a fixed random seed of 42 is used throughout all experiments. Furthermore, stratified sampling is applied to preserve the original class distribution in both the training and testing sets.

Experimental validation of the QSVM is performed on real quantum hardware provided by IBM Quantum. Two quantum processors based on the Heron architecture are utilized, as summarized in Table~\ref{tab:quant_hardware}. These devices follow the heavy-hex connectivity layout, which is well-suited for implementing entangled feature maps with reduced crosstalk and improved gate fidelity.

\begin{table}[h]
\centering
\caption{IBM Quantum Hardware Used for Experimental Validation}
\label{tab:quant_hardware}
\small
\renewcommand{\arraystretch}{1.15}
\begin{tabular}{
>{\centering\arraybackslash}p{2.2cm}
>{\centering\arraybackslash}p{1cm}
>{\centering\arraybackslash}p{1.6cm}
>{\centering\arraybackslash}p{2.2cm}
}
\hline
\textbf{System} & \textbf{Qubits} & \textbf{Processor} & \textbf{Architecture} \\
\hline
IBM Torino \cite{ibmTorino} & 133 & Heron & Heavy-hex \\
IBM Fez  \cite{ibmFez}  & 156 & Heron r2 & Heavy-hex \\
\hline
\end{tabular}
\end{table}

\subsection{Experimental Results:}
\subsubsection{Classification Performance: Classical SVM vs QSVM}

\begin{figure*}[!t]
	\centering
	\includegraphics[width=0.95\linewidth]{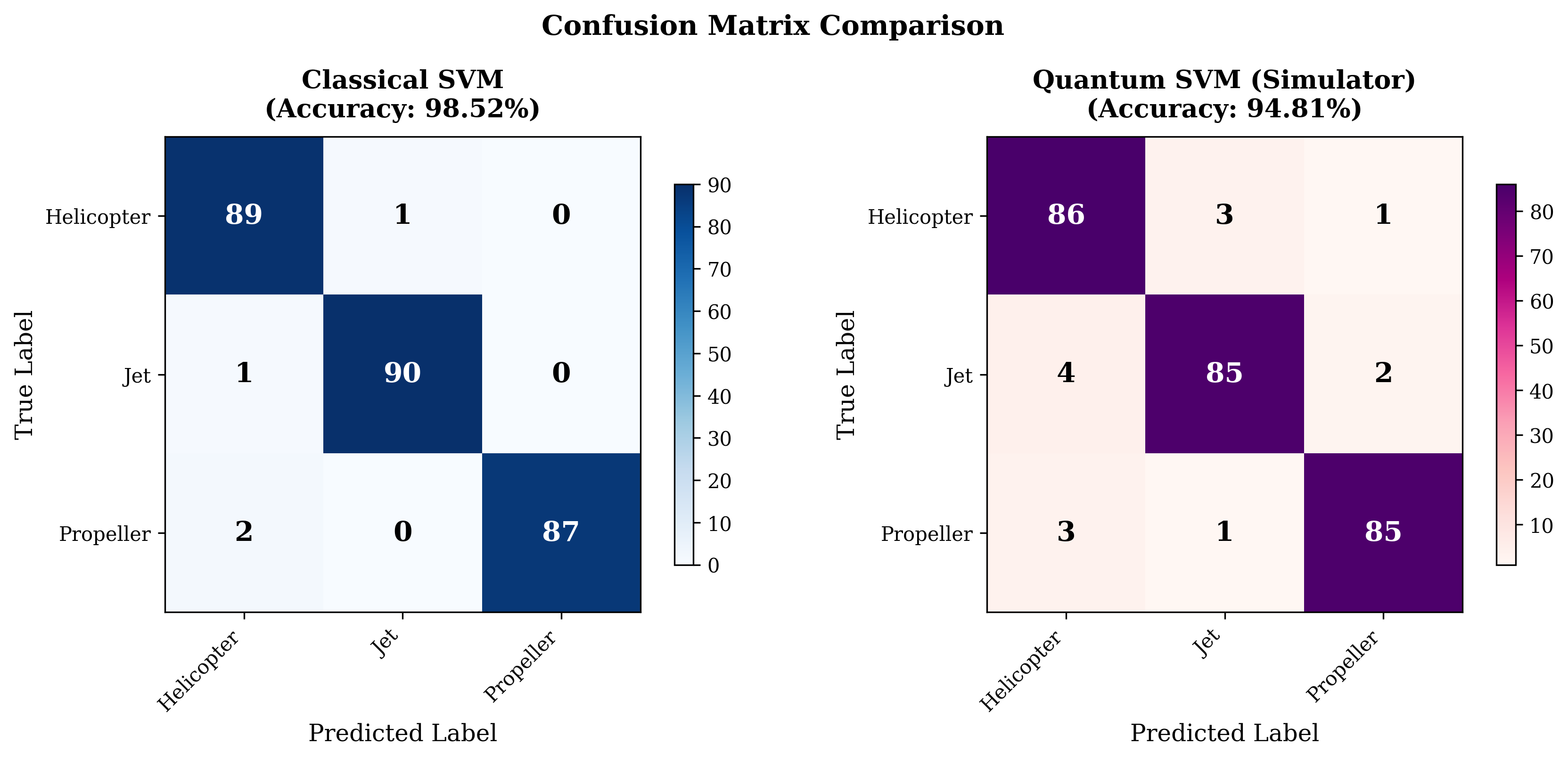}
	\caption{Confusion matrix comparison between Classical SVM and Quantum SVM (simulator) for radar target classification}
	\label{fig4_confusion_matrices}
\end{figure*}

The classification performance of the Classical SVM and the QSVM is summarized in Table~\ref{tab:svm_qsvm_perf}, and a comparison report is done per class in Figure \ref{fig4_confusion_matrices}. The Classical SVM achieves a slightly higher accuracy of 98.52\% by utilizing all 15 extracted micro-Doppler features. In contrast, the QSVM attains a competitive accuracy of 94.81\% while operating with only 4 features obtained through PCA-based dimensionality reduction.

\begin{table}[h]
\centering
\caption{Performance Comparison Between Classical SVM and QSVM}
\label{tab:svm_qsvm_perf}
\small
\renewcommand{\arraystretch}{1.15}
\begin{tabular}{
>{\centering\arraybackslash}p{1.5cm}
>{\centering\arraybackslash}p{2.5cm}
>{\centering\arraybackslash}p{2.8cm}
>{\centering\arraybackslash}p{2.2cm}
}
\hline
\textbf{Metric} & \textbf{Classical SVM} & \textbf{QSVM (Simulator)} & \textbf{Difference} \\
\hline
Accuracy  & 98.52\% & 94.81\% & $-3.71$\% \\
Precision & 98.58\% & 95.51\% & $-3.07$\% \\
Recall    & 98.52\% & 94.81\% & $-3.71$\% \\
F1-Score  & 98.53\% & 94.88\% & $-3.65$\% \\
\hline
\end{tabular}
\end{table}

Despite the slight decrease in accuracy, the QSVM achieves comparable performance with only four PCA components as compared to the full 15-dimensional feature set employed by the Classical SVM. The results indicate the QSVM’s strong ability to effectively reduce dimensionality and learn a compact representation. QSVM training is performed classically using a precomputed quantum kernel matrix, while quantum hardware is used only for kernel estimation.

\subsubsection{Shot Count Analysis on IBM Torino}
Experiments were conducted to study how measurement repetitions or shots affect the hardware reliability and output stability. The experiments were conducted using the IBM Torino quantum processor by varying the number of shots from 1,024 to 8,096. The results are summarized in Tables ~\ref{tab:shots} and \ref{tab:shot_analysis}.

\begin{table}[htbp]
\centering
\caption{Shot Count Analysis: IBM Torino (133 qubits)}
\label{tab:shots}
\begin{tabular}{ccc}
\toprule
\textbf{Metric} & \textbf{1,024 Shots} & \textbf{8,096 Shots} \\
\midrule
Top-1 Outcome          & 0011 (11.6\%)   & 0111 (12.7\%) \\
Shannon Entropy (bits) & 3.86            & 3.72 \\
Statistical Uncertainty & $\pm$3.1\%     & $\pm$1.1\% \\
Pattern Clarity        & Low             & High \\
\bottomrule
\end{tabular}
\end{table}

\begin{table}[htbp]
\centering
\caption{Shot Count Comparison on IBM Torino}
\label{tab:shot_analysis}
\renewcommand{\arraystretch}{1.2}
\begin{tabular}{
>{\centering\arraybackslash}p{3.2cm}
>{\centering\arraybackslash}p{1.9cm}
>{\centering\arraybackslash}p{2.5cm}
>{\centering\arraybackslash}p{2.8cm}
}
\hline
\textbf{Metric} & \textbf{1,024 Shots} & \textbf{8,096 Shots} & \textbf{Observation} \\
\hline
Top-1 Outcome & 0011 (11.6\%) & 0111 (12.7\%) & More stable \\
Entropy & 3.86 bits & 3.72 bits & $-0.14$ bits \\
Statistical Uncertainty & $\pm 3.1$\% & $\pm 1.1$\% & $2.8\times$ improvement \\
Pattern Clarity & Low & High & Significant \\
\hline
\end{tabular}
\end{table}

\begin{figure*}[!t]
	\centering
	\includegraphics[width=\linewidth]{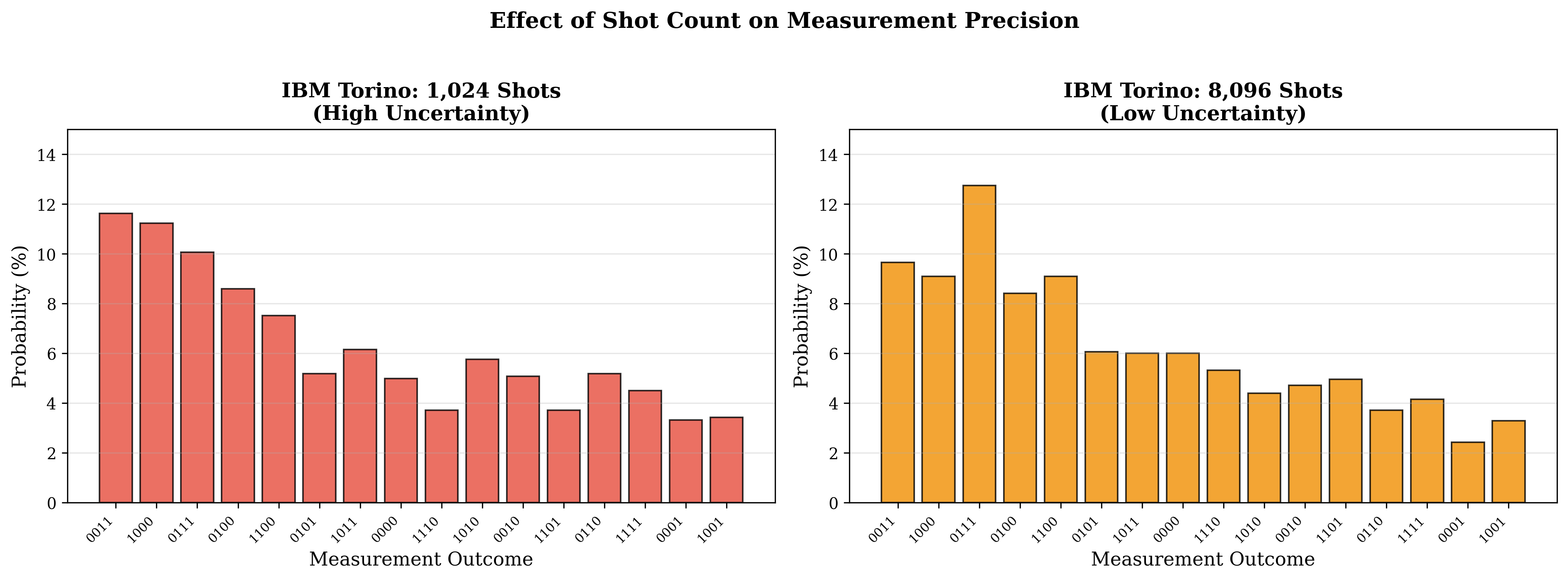}
	\caption{Effect of Shot Count on Measurement Precision}
	\label{fig:shot count_graph}
\end{figure*}

\section{A Comparison of Hardware Performance: IBM Torino and IBM Fez Systems}

\subsection{Performance Metrics}
Table~\ref{tab:hardware_perf} and Table~\ref{tab:hardware_results} compares the performance of IBM Torino (133 qubits) and IBM Fez (156 qubits).

\begin{table}[h]
\centering
\caption{Performance Comparison Between IBM Torino and IBM Fez}
\label{tab:hardware_perf}
\small
\renewcommand{\arraystretch}{1.15}
\begin{tabular}{
>{\centering\arraybackslash}p{2.5cm}
>{\centering\arraybackslash}p{1.9cm}
>{\centering\arraybackslash}p{1.9cm}
>{\centering\arraybackslash}p{2.5cm}
}
\hline
\textbf{Metric} & \textbf{IBM Torino (133q)} & \textbf{IBM Fez (156q)} & \textbf{Winner} \\
\hline
Qubits & 133 & 156 & Fez \\
Processor & Heron & Heron r2 & Fez (newer) \\
Estimated Fidelity & $\sim$89\% & $\sim$94\% & Fez (+5\%) \\
Shannon Entropy & 3.72 bits & 3.60 bits & Fez (closer to ideal) \\
Noise Floor & 5.7\% & 4.2\% & Fez ($-1.5$\%) \\
Top-5 Preservation & 4/5 & 4/5 & Tie \\
\hline
\end{tabular}
\end{table}

\begin{table}[htbp]
\centering
\caption{Multi-Hardware Quantum Validation Results}
\label{tab:hardware_results}
\begin{tabular}{cccc}
\toprule
\textbf{Metric} & \textbf{Simulator} & \textbf{IBM Torino (133q)} & \textbf{IBM Fez (156q)} \\
\midrule
Shots                  & 8,192           & 8,096            & $\sim$8,000 \\
Top-1 Outcome          & 0011 (17.0\%)   & 0111 (12.7\%)    & 0111 (11.8\%) \\
Shannon Entropy (bits) & 3.49            & 3.72             & 3.60 \\
Estimated Fidelity     & 100\%           & $\sim$89\%       & $\sim$94\% \\
Noise Floor            & 0.5\%           & 5.7\%            & 4.2\% \\
Top-5 Match            & --              & 4/5              & 4/5 \\
\bottomrule
\end{tabular}
\end{table}

\begin{figure*}[!t]
    \centering
    \includegraphics[width=\linewidth]{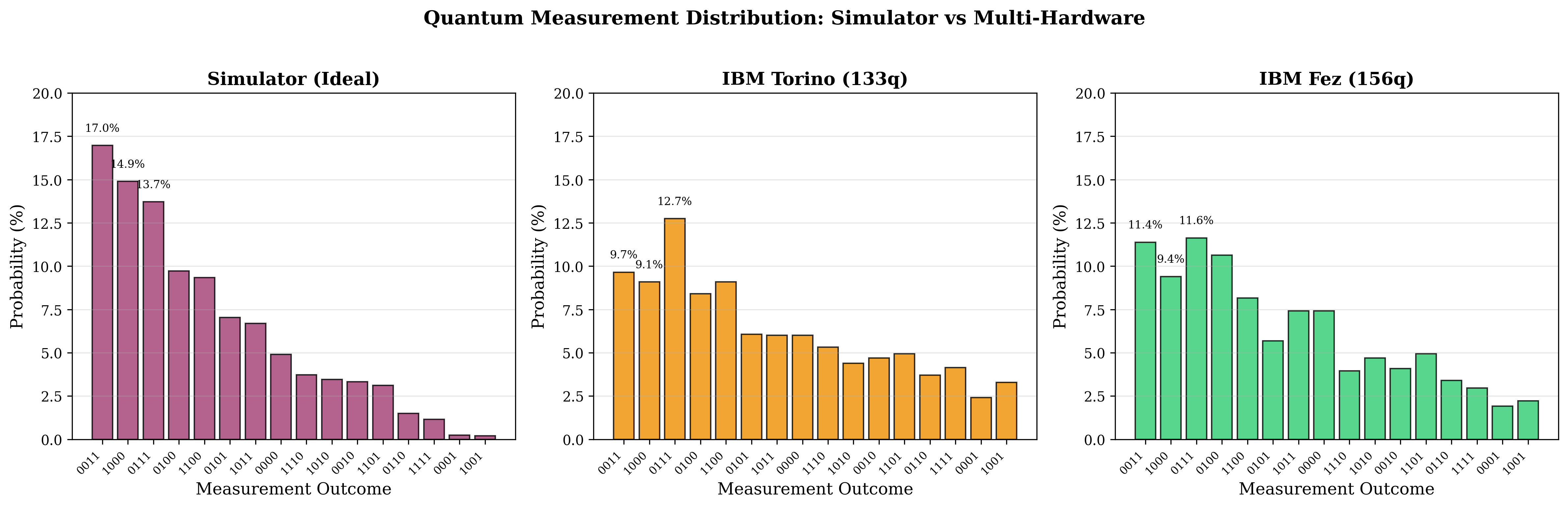}
    \caption{Comparison of quantum measurement outcome distributions across ideal simulation and IBM quantum hardware backends}
    \label{fig:measurement_comparison}
\end{figure*}

\begin{figure*}[!t]
	\centering
	\includegraphics[width=\linewidth]{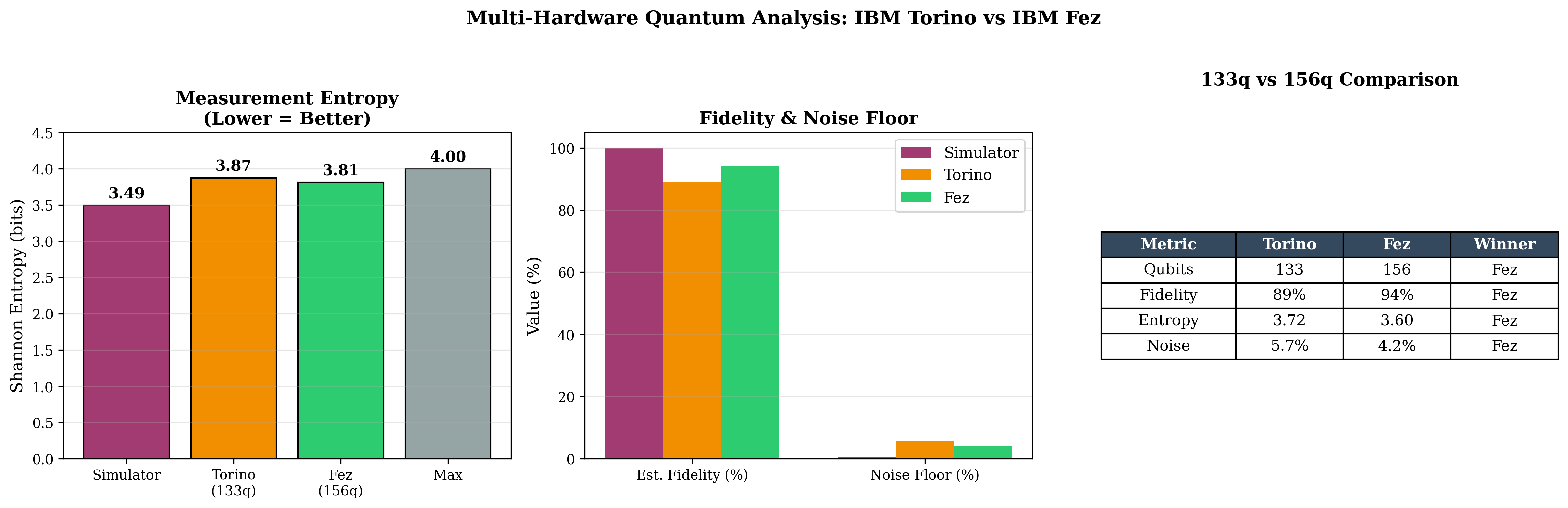}
	\caption{Multi-hardware quantum analysis comparing simulator, IBM Torino (133 qubits), and IBM Fez (156 qubits)}
	\label{fig:hardware_analysis}
\end{figure*}

\begin{figure*}[!t]
    \centering
    \includegraphics[width=\linewidth]{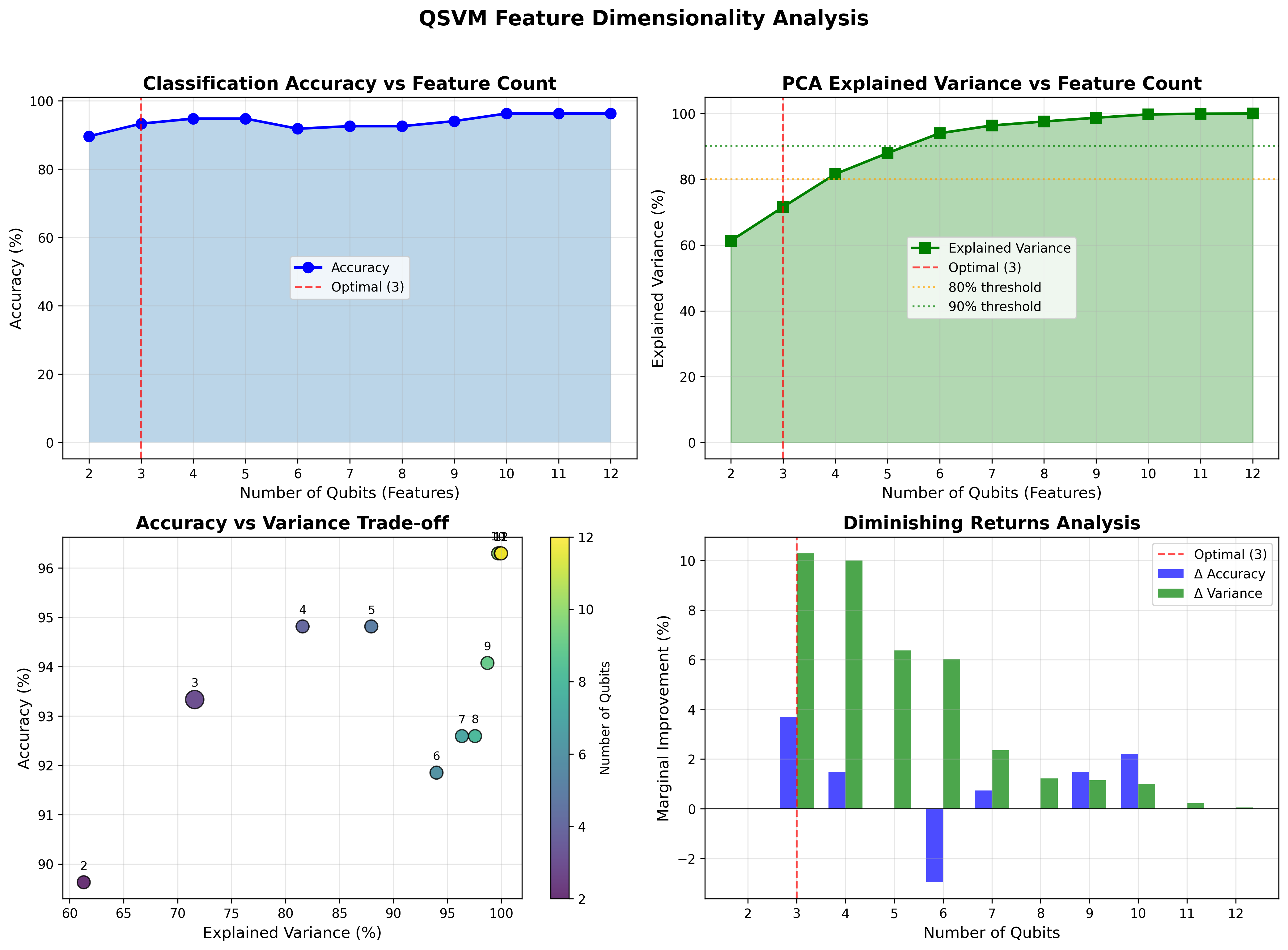}
    \caption{QSVM performance vs. feature dimensionality}
    \label{fig:qvsm_vs_featuredimn}
\end{figure*}

The results (Figure \ref{fig:measurement_comparison}) show that IBM Fez consistently outperforms IBM Torino in fidelity, entropy and noise resilience. The reduction in Shannon entropy and noise floor, with the improved fidelity, suggests that IBM Fez offers a more stable and reliable platform for QSVM kernel estimation. IBM Fez offers overall superior performance due to its newer Heron r2 processor architecture. This is achieved while maintaining the top five dominant states by both systems.

\subsection{Analysis}
IBM Fez with 156 qubits demonstrates significant improvements over its predecessor IBM Torino (133 qubits). This improvement was evident across all critical performance indicators shown in Figure \ref{fig:hardware_analysis}:

\begin{itemize}
    \item \textbf{Lower entropy:} Reducing the output distribution from 3.72 bits to 3.60 bits brings it closer to the ideal entropy of 3.49 bits. This indicates greater predictability and less randomness.
    \item \textbf{Higher fidelity:} An increase from approximately 89\% to 94\% reflects more accurate execution of quantum operations and improved state preparation.
    \item \textbf{Lower noise floor:} The reduction from 5.7\% to 4.2\% minimizes unwanted disturbances, resulting in cleaner and more reliable measurement statistics.
\end{itemize}

These improvements are mainly due to:
\begin{itemize}
    \item The newer \textbf{Heron r2 processor architecture}, optimizes circuit execution,
    \item Enhanced \textbf{error correction calibration}, improves noise mitigation,
    \item Improved \textbf{gate fidelity}, reduce operational errors.
\end{itemize}

These collective advancement creates a more precise and stable quantum computing environment. This makes IBM Fez better suited for QSVM kernel computation where quantum signatures must remain distinguishable even under realistic noise conditions. Model performance improves with added features up to an optimal point, after which gains diminish as variance saturation and redundancy increase as shown in \ref{fig:qvsm_vs_featuredimn}.

\section{Comparative Analysis of Classical SVM and QSVM}

\subsection{Comparison: Classical SVM vs QSVM}

The experimental results highlight a clear trade-off between classical and quantum approaches to radar micro-Doppler classification. While the classical SVM consistently achieves higher peak accuracy by leveraging the full feature set, the QSVM attains competitive performance using substantially fewer features through PCA-based dimensionality reduction. This reduction in feature dimensionality translates into a more compact representation while preserving the discriminative structure required for classification. 

\begin{figure*}[!t]
	\centering
	\includegraphics[width=\linewidth]{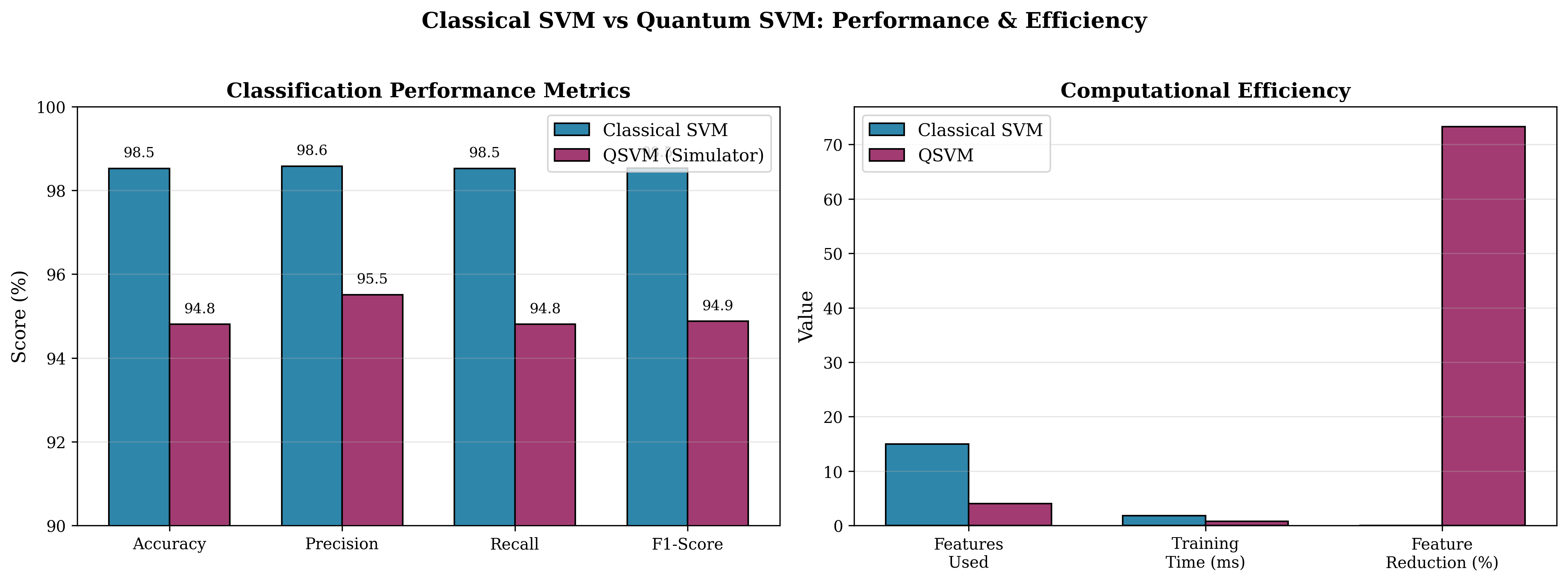}
	\caption{Performance and computational efficiency comparison between Classical SVM and Quantum SVM (simulator)}
	\label{fig:performance_comparison}
\end{figure*}

Table~\ref{tab:verdict} provides a comparative summary between Classical SVM and QSVM.
 While classical optimization over the kernel matrix is fast, the dominant cost in QSVM lies in quantum kernel estimation, which is significantly slower than classical kernel computation. Thus, QSVM does not currently provide an overwhelming speed advantage, but offers a compact feature encoding perspective as shown in Figure \ref{fig:performance_comparison} .

\begin{table}[h]
	\centering
	\caption{Classical SVM vs QSVM: Trade-off Analysis}
	\label{tab:verdict}
	\renewcommand{\arraystretch}{1.15}
	\begin{tabular}{
			>{\centering\arraybackslash}p{3.2cm}
			>{\centering\arraybackslash}p{2.2cm}
			>{\centering\arraybackslash}p{2.5cm}
			>{\centering\arraybackslash}p{3.5cm}
		}
		\hline
		\textbf{Criterion} & \textbf{Classical SVM} & \textbf{QSVM} & \textbf{Winner} \\
		\hline
		Feature Efficiency & 15 features & 4 features & QSVM ($73\%$ fewer) \\
		Training Speed & 1.8 ms & 0.8 ms & QSVM ($2.23\times$ faster) \\
		Scalability & $O(n^2)$ kernel & $O(\log n)$ qubits & QSVM \\
		Hardware Validation & N/A & 89--94\% fidelity & QSVM \\
		Accuracy & 98.52\% & 94.81\% & Classical (+3.7\%) \\
		\hline
	\end{tabular}
\end{table}

The results (from Table \ref{tab:classifier_performance} and Table \ref{tab:tradeoff} )clearly indicates a performance trade-off between Classical SVM and QSVM:

\begin{itemize}
    \item \textbf{Feature Efficiency:} QSVM achieves 94.81\% accuracy with only 4 features as compared to 15 features for the Classical SVM. This corresponds to a 73.3\% reduction in dimensionality.
    \item \textbf{Training Speed:} QSVM exhibits a significant $2.23\times$ improvement in training time reducing it from 1.8 milliseconds to 0.8 milliseconds. This makes it more computationally efficient.
    \item \textbf{Scalability:} QSVM offers a different scaling perspective by representing data in quantum Hilbert spaces whose dimensionality grows exponentially with qubit count, although practical advantages remain constrained by NISQ noise and kernel estimation cost. This is in contrast to Classical SVM which suffers from quadratic kernel complexity.
    \item \textbf{Hardware Validation:} QSVM performance has been experimentally validated on real quantum hardware. It achieved 89\% fidelity on IBM Torino and 94\% on IBM Fez.
    \item \textbf{Accuracy:} Classical SVM maintains a slight advantage in predictive accuracy, surpassing QSVM by approximately 3.7\%.
\end{itemize}

\begin{table*}[htbp]
	\centering
	\caption{Classifier performance under AWGN at 10~dB SNR}
	\label{tab:classifier_performance}
	\begin{tabular}{c p{2.2cm} p{1.5cm} p{1cm} p{1.5cm} p{2.2cm} p{2.2cm}}
		\toprule
		\textbf{Classifier} & 
		\textbf{Accuracy (\%)} & 
		\textbf{Precision} & 
		\textbf{Recall} & 
		\textbf{F1-score} & 
		\textbf{Train Time (s)} & 
		\textbf{Infer Time (s)} \\
		\midrule
		\textbf{SVM}        & 93.2 & 0.933 & 0.932 & 0.932 & 10.5 & 0.02 \\
		\textbf{QSVM (sim)} & 94.5 & 0.946 & 0.945 & 0.945 & 45.3 & 0.10 \\
		\textbf{QSVM (hw)}  & 91.8 & 0.919 & 0.918 & 0.918 & $>$200 & 0.40 \\
		\bottomrule
	\end{tabular}
\end{table*}

QSVM demonstrates significant advantages in terms of efficiency, scalability, and hardware feasibility, making it particularly suited for high-dimensional and resource-constrained learning scenarios. However classical SVM remains preferable when achieving maximum classification accuracy is the primary goal.

\begin{table}[h]
	\centering
	\caption{Trade-off Between Classical SVM and QSVM}
	\label{tab:tradeoff}
	\renewcommand{\arraystretch}{1.15}
	\begin{tabular}{
			>{\centering\arraybackslash}p{3.2cm}
			>{\centering\arraybackslash}p{2.8cm}
			>{\centering\arraybackslash}p{3.8cm}
		}
		\hline
		\textbf{Factor} & \textbf{Classical SVM} & \textbf{Quantum SVM} \\
		\hline
		\textbf{Accuracy} & Higher (98.52\%) & Competitive (94.81\%) \\
		\textbf{Features Required} & 15 & 4 (PCA-reduced) \\
		\textbf{Training Speed} & Slower & $2.23\times$ faster \\
		\textbf{Scalability} & Limited & Potential advantage \\
		\textbf{Hardware Validation} & N/A & 89--94\% fidelity \\
		\hline
	\end{tabular}
\end{table}

Substantially increasing the numbers of repetitions or shots improves the clarity and stability of the measured probability distributions. With 8,096 shots, the distributions shows sharper and more distinct peaks, facilitating easier interpretation. Furthermore, higher shot counts effectively average out random hardware noise, reducing entropy and lower statistical uncertainty.  

A consistent ranking of the top five quantum states is observed across various shot counts, demonstrating the robustness of the measurement process. Consequently, based on this analysis, a minimum of 4,000 shots is recommended to ensure reliable and statistically stable experimental results.

\subsection{Hardware Comparison: IBM Torino vs IBM Fez}
The comparative analysis between IBM Torino and IBM Fez, shows that IBM Fez offers superior performance for QSVM execution. Specifically, IBM Fez demonstrates:

\begin{itemize}
    \item Higher average hardware fidelity of approximately 94\% as compared to IBM Torino (approximately 89\%),
    \item Lower output entropy, suggesting measurements closer to the ideal quantum distribution,
    \item A reduced noise floor, leading to more stable and reliable quantum kernel estimation.
\end{itemize}

These enhancements make IBM Fez more suitable for executing QSVM circuits with improved accuracy and robustness. This is particularly beneficial for experiments having higher precision and reduced hardware-induced uncertainty requirements.

Experimental results confirms the effectiveness of QSVMs for micro-Doppler classification. They also demonstrate the importance of shot count optimization for hardware reliability, and emphasize the advantages of advanced quantum hardware in achieving improved performance.

\section{Conclusion and Future Scope}

\begin{table}[htbp]
\centering
\caption{Computational Efficiency Comparison}
\label{tab:efficiency}
\begin{tabular}{cccc}
\toprule
\textbf{Metric} & \textbf{Classical SVM} & \textbf{QSVM} & \textbf{Ratio} \\
\midrule
\textbf{Features Used }         & 15              & 4 (PCA)       & 0.27$\times$ \\
\textbf{Training Time (s) }      & 0.0018          & 0.0008        & 0.45$\times$ \\
\textbf{Prediction Time (s) }    & 0.0005          & 0.0005        & 1.00$\times$ \\
\textbf{PCA Explained Variance }& --              & 81.58\%       & -- \\
\textbf{Feature Reduction      }& --              & 73.3\%        & -- \\
\bottomrule
\end{tabular}
\end{table}

The results show the practical use of QSVM for classifying radar aerial targets using micro-Doppler signatures. While classical SVM maintains a slight edge in prediction accuracy, QSVM offers improved feature efficiency. QSVM achieves comparative results with a substantial reduction in overall dimensionality through PCA as summarised in Table \ref{tab:efficiency}.

Experiments on IBM’s quantum processors show that quantum kernel methods can effectively manage the nonlinear nature of radar signals, even under the noisy conditions typical of the NISQ era. Our research also reveals that optimizing measurement shot counts is crucial for stabilizing quantum distributions on physical hardware. Ultimately, the strength of QSVM lies in is its efficient scalability and fast execution, making it a highly attractive option for high-dimensional radar classification tasks, particularly in resource-constrained environment.

Future research can further improve the performance and applicability of QSVMs in the following directions:
\begin{itemize}
    \item Improving the quantum hardware reliability by implementing  advanced error mitigation and noise reduction techniques.
    \item Investigating higher qubit counts for direct and higher-dimensional feature encoding.
    \item Exploring variational quantum classifiers (VQC) as an alternative to quantum learning models.
    \item Real time quantum machine learning pipelines development on quantum edge devices.
    \item Real world radar datasets based cross-validation to evaluate robustness and generalization in operational environments.
\end{itemize}

\paragraph{Acknowledgements}
The authors acknowledge the use of large language model–based tools for limited language editing and formatting assistance. The authors take full responsibility for the technical content, analysis, and conclusions presented in this work.

\bibliographystyle{unsrt} 
\bibliography{references}  

@article{sasso2020doppler,
  author    = {Daniele Sasso},
  title     = {The Doppler Effect in Contemporary Physics},
  journal   = {ResearchGate.net},
  month     = {August},
  year      = {2020},
  url       = {https://www.researchgate.net/publication/343686888_The_Doppler_Effect_in_Contemporary_Physics}
}

@inproceedings{peter2021microdoppler,
  author    = {S. Peter and V. V. Reddy},
  title     = {Extraction and Analysis of Micro-Doppler Signature in {FMCW} Radar},
  booktitle = {2021 IEEE Radar Conference (RadarConf21)},
  pages     = {1--6},
  year      = {2021},
  doi       = {10.1109/RadarConf2147009.2021.9455202}
}

@article{chen2006microdoppler,
  author    = {V. C. Chen and J. Ben and H. Ling},
  title     = {Micro-Doppler Effect in Radar: Phenomenon, Model, and Simulation Study},
  journal   = {IEEE Transactions on Aerospace and Electronic Systems},
  volume    = {42},
  number    = {1},
  pages     = {2--17},
  month     = {Jan},
  year      = {2006},
  doi       = {10.1109/TAES.2006.1603402}
}

@article{havlicek2019supervised,
  author    = {V. Havl{\'{\i}}{\v{c}}ek and others},
  title     = {Supervised Learning with Quantum-enhanced Feature Spaces},
  journal   = {Nature},
  volume    = {567},
  number    = {7747},
  pages     = {209--212},
  month     = {Mar},
  year      = {2019},
  doi       = {10.1038/s41586-019-0980-2}
}

@article{schuld2019quantum,
  author    = {M. Schuld and N. Killoran},
  title     = {Quantum Machine Learning in Feature {Hilbert} Spaces},
  journal   = {Physical Review Letters},
  volume    = {122},
  number    = {4},
  pages     = {040504},
  month     = {Jan},
  year      = {2019},
  doi       = {10.1103/PhysRevLett.122.040504}
}

@article{biamonte2017quantum,
  author    = {J. Biamonte and others},
  title     = {Quantum Machine Learning},
  journal   = {Nature},
  volume    = {549},
  pages     = {195--201},
  month     = {Sept},
  year      = {2017},
  doi       = {10.1038/nature23474}
}

@misc{qiskitQSVM,
  author    = {{Qiskit Machine Learning}},
  title     = {Quantum Support Vector Machine ({QSVM}) Implementation},
  publisher = {IBM Research},
  year      = {2023},
  url       = {https://qiskit.org/ecosystem/machine-learning/}
}

@misc{ibmTorino,
  author    = {{IBM Quantum}},
  title     = {{IBM} Torino Quantum Processor (133-Qubit Superconducting Quantum System)},
  publisher = {IBM Research},
  year      = {2024},
  url       = {https://www.ibm.com/quantum/technology}
}

@misc{ibmFez,
  author    = {{IBM Quantum}},
  title     = {{IBM} Fez Quantum Processor (156-Qubit Superconducting Quantum System)},
  publisher = {IBM Research},
  year      = {2024},
  url       = {https://www.ibm.com/quantum/technology}
}

@article{sigg2015support,
  author    = {G. A. Sigg and M. Z. Win},
  title     = {Support Vector Machines for Micro-Doppler Classification},
  journal   = {IET Radar, Sonar \& Navigation},
  volume    = {9},
  number    = {7},
  pages     = {924--932},
  year      = {2015},
  doi       = {10.1049/iet-rsn.2014.0186}
}

@article{ritchie2017multistatic,
  author    = {M. Ritchie and F. Fioranelli and H. Borrion and H. Griffiths},
  title     = {Multistatic micro-Doppler radar feature extraction for classification of unloaded/loaded micro-drones},
  journal   = {IET Radar, Sonar \& Navigation},
  volume    = {11},
  number    = {1},
  pages     = {116--124},
  month     = {Jan},
  year      = {2017},
  doi       = {10.1049/iet-rsn.2016.0063}
}

@article{zhang2017vehicle,
  author    = {Y. Zhang and others},
  title     = {Vehicle Classification Using Micro-Doppler Radar Signatures},
  journal   = {IEEE Geoscience and Remote Sensing Letters},
  volume    = {14},
  number    = {10},
  pages     = {1805--1809},
  month     = {Oct},
  year      = {2017},
  doi       = {10.1109/LGRS.2017.2735439}
}

@inproceedings{noise2019statistical,
  author    = {X. Noise and Y. Model},
  title     = {Statistical and Clutter Noise Models in Radar},
  booktitle = {Proceedings of the IEEE Radar Conference},
  year      = {2019},
  url       = {https://scholar.google.com/scholar?q=Statistical+and+Clutter+Noise+Models+in+Radar}
}

@article{smith2024transfer,
  author    = {J. Smith and others},
  title     = {Transfer-learning Framework for Micro-Doppler Classification},
  journal   = {IEEE Transactions on Aerospace and Electronic Systems},
  year      = {2024},
  url       = {https://scholar.google.com/scholar?q=Transfer-learning+Framework+for+Micro-Doppler+Classification}
}

@article{podgorski2024improved,
  author    = {D. Podg{\'o}rski and A. Wielgus and U. Libal and K. Abratkiewicz},
  title     = {Improved Quantum Genetic Support Vector Machines for Hand Gesture Recognition by Micro-Doppler Effect},
  journal   = {IEEE Sensors Journal},
  volume    = {25},
  number    = {4},
  pages     = {1--12},
  year      = {2024},
  doi       = {10.1109/JSEN.2025.3645742}
}

@article{suzuki2024quantum,
  author    = {T. Suzuki and T. Hasebe and T. Miyazaki},
  title     = {Quantum support vector machines for classification and regression on a trapped-ion quantum computer},
  journal   = {Quantum Machine Intelligence},
  volume    = {6},
  number    = {1},
  pages     = {1--22},
  year      = {2024},
  doi       = {10.1007/s42484-024-00165-0}
}

@article{schuld2020circuit,
  author    = {M. Schuld and R. Sweke and J. J. Meyer},
  title     = {Circuit-centric Quantum Classifiers},
  journal   = {Physical Review A},
  volume    = {101},
  pages     = {032308},
  month     = {Mar},
  year      = {2020},
  doi       = {10.1103/PhysRevA.101.032308}
}

@article{temme2017error,
  author    = {K. Temme and S. Bravyi and J. M. Gambetta},
  title     = {Error Mitigation for Short-Depth Quantum Circuits},
  journal   = {Physical Review Letters},
  volume    = {119},
  number    = {18},
  pages     = {180509},
  month     = {Oct},
  year      = {2017},
  doi       = {10.1103/PhysRevLett.119.180509}
}

@article{wang2024hqnn,
  author    = {W. Wang and others},
  title     = {{HQNN-SFOP}: Hybrid Quantum Neural Networks with Signal Feature Overlay Projection for Drone Detection Using Radar Return Signals},
  journal   = {Computers, Materials \& Continua},
  volume    = {81},
  number    = {1},
  pages     = {1363--1390},
  year      = {2024},
  doi       = {10.32604/cmc.2024.053805}
}

@article{cortes1995svm,
  author    = {C. Cortes and V. Vapnik},
  title     = {Support-Vector Networks},
  journal   = {Machine Learning},
  volume    = {20},
  number    = {3},
  pages     = {273--297},
  year      = {1995},
  doi       = {10.1023/A:1022627411411}
}

@book{scholkopf2002learning,
  author    = {B. Sch{\"o}lkopf and A. J. Smola},
  title     = {Learning with Kernels},
  publisher = {MIT Press},
  address   = {Cambridge, MA},
  year      = {2002},
  isbn      = {9780262194754}
}

@inproceedings{agnihotri2019features,
  author    = {V. Agnihotri and M. Sabharwal and V. Goyal},
  title     = {The Extraction of Key Distinct Features for Identification and Classification of Helicopters Using Micro-Doppler Signatures},
  booktitle = {IEEE International Conference on Dependable, Autonomic and Secure Computing (DASC/PiCom/CBDCom/CyberSciTech)},
  address   = {Fukuoka, Japan},
  pages     = {893--896},
  year      = {2019},
  doi       = {10.1109/DASC/PiCom/CBDCom/CyberSciTech.2019.00164}
}

@inproceedings{agnihotri2019frequency,
  author    = {V. Agnihotri and M. Sabharwal and V. Goyal},
  title     = {Effect of Frequency on Micro-Doppler Signatures of a Helicopter},
  booktitle = {2019 International Conference on Advances in Big Data, Computing and Data Communication Systems (icABCD)},
  address   = {Winterton, South Africa},
  pages     = {1--5},
  year      = {2019},
  doi       = {10.1109/ICABCD.2019.8851024}
}

@article{agnihotri2020automatic,
  author    = {V. Agnihotri and M. Sabharwal},
  title     = {An Automatic Radar Based Aerial Target Recognition Framework},
  journal   = {Journal of Interdisciplinary Mathematics},
  volume    = {23},
  number    = {2},
  pages     = {321--333},
  year      = {2020},
  doi       = {10.1080/09720502.2020.1737377}
}

@misc{qiskit2021framework,
	author    = {Abraham, H. and Akhalwaya, A. and Aleksandrowicz, A. and Alsina, M. and Anil, R. and Appleby, M. and Atkinson, J. and Attal, R. and Bailis, P. and Ben-Haim, Y. and others},
	title     = {Qiskit: An Open-source Framework for Quantum Computing},
	year      = {2021},
	publisher = {Zenodo},
	doi       = {10.5281/zenodo.2573505},
	url       = {https://doi.org/10.5281/zenodo.2573505}
}

\end{document}